\documentclass[print]{aa}  

\usepackage{graphicx}

\usepackage{siunitx}

\usepackage{txfonts}

\newcommand{\bb}[1]{\boldsymbol#1}

\begin{document}

   \title{Scattering, absorption, and thermal emission by large cometary dust particles}

   \subtitle{Synoptic numerical solution}

\titlerunning{Scattering, absorption and emission by cometary dust}

   \author{Johannes Markkanen\inst{1}
          \and
          Jessica Agarwal\inst{1}
          }
   \institute{Max Planck Institute for Solar System Research, Justus-von-Liebig-Weg 3, 37077 G\"ottingen, Germany\\
\email{markkanen@mps.mpg.de}}
 \date{}

\authorrunning{J. Markkanen and J. Agarwal}

 
  \abstract
   {Remote light scattering and thermal infrared observations provide  clues about the physical properties of cometary and interplanetary dust particles. Identifying these properties will lead to a better understanding of the formation and evolution of the Solar System.}
   {We present a numerical solution for the radiative and conductive heat transport in a random particulate medium enclosed by an arbitrarily shaped surface. The method will be applied to study thermal properties of cometary dust particles.}
   {The recently introduced incoherent Monte Carlo radiative transfer method developed for scattering, absorption, and propagation of electromagnetic waves in dense discrete random media is extended for radiative heat transfer and thermal emission. The solution is coupled with the conductive Fourier transport equation that is solved with the finite-element method.}
   {The proposed method allows the synoptic analysis of light scattering and thermal emission by large cometary dust particles consisting of submicrometer-sized grains. In particular, we show that these particles can sustain significant temperature gradients resulting in the superheating factor phase function observed for the coma of comet 67P/Churyumov-Gerasimenko. }
   {}


   \maketitle
%

\section{Introduction}

The physical properties of cosmic dust particles, which include size, shape, porosity, and composition, can be constrained from remote light  scattering and thermal emission observations. Observations of scattered solar and thermally emitted light have mostly  been interpreted separately using different particle models and numerical methods \citep{Kolokolova2004, Kimura2016}. This may lead to contradictions in the retrieved particle properties as the interpretation procedure is not self-consistent. Using the same numerical method and the same particle model for scattered solar and thermally emitted light (i.e.,  {\em \emph{synoptic modeling)}} results in more reliable estimates for the physical properties of dust particles. Such modeling, however, introduces various numerical challenges due to the multiscale nature of the problem. The solution is needed from ultraviolet to thermal infrared. In addition, cometary dust particles contain multiscale structures. They are thought to be large aggregates consisting of submicrometer-sized grains \citep{Guettler2019}.  

Numerically exact electromagnetic techniques allow for analyses of scattering and radiative heat transfer properties of small aggregated particles. Combining the standard electromagnetic methods with the statistical fluctuation driven electrodynamic solvers, the radiative heat transfer problem can be solved directly from Maxwell's equations \citep{Jin2017}. This approach is computationally heavy, and consequently it is not applicable to particles that are much larger than the optical wavelengths or that contain a large number of individual grains. For sufficiently large  objects such as asteroids and comet nuclei, a typical approach is to localize absorption and emission sources on the surfaces of the object using a proper boundary condition. Then, by introducing the effective conductive and radiative heat transfer coefficient, heat transport inside the object can be modeled \citep{huebner2006, delbo2015asteroid}. This  simple model is a valid approximation only for one-dimensional problems, and it introduces significant errors when the object has locally three-dimensional features (on length scales on the order of the mean free path length). Thus, large objects with small details cannot be treated rigorously with the existing computational techniques, due to the large computational time and memory requirement or inadequate approximations.

It is also possible to use the radiative transfer equation (RTE) in combination with the Fourier heat transport equation to deal with heat transfer in participating media \citep{Howell2010,Modest2013_10}. A large number of different solution techniques to solve the RTE exist, for example  the discrete ordinates \citep{Hardy2016}, finite-volume \citep{Coelho2014}, finite-element methods \citep{Richling2001}, and Monte Carlo ray tracing methods  \citep{Howell1998}. Nonetheless, the standard RTE is only applicable to sparse discrete random media where the grains are in each other's far zone. Recently, a numerical framework for electromagnetic scattering, absorption, and propagation in dense discrete random media was introduced which extends the applicability range of the standard RTE \citep{Muinonen2018, Markkanen2018, Vaisanen2019}. Furthermore, \cite{Markkanen2018_2} augmented the method with the coherent field contribution to account for surface reflections and refractions allowing the  scattering analysis of millimeter-sized cometary dust particles consisting of submicrometer-sized grains. 

In this paper we extend the numerical method developed in \cite{Markkanen2018_2} for the  radiative heat transfer problems. The radiative heat transfer solution is coupled with the transient conductive heat transfer equation, which is solved by the finite-element method (FEM). The presented approach allows us to analyze the scattering and thermal properties of particles consisting of aggregates of small grains over a wide wavelength range. Finally, we show that large cometary dust particles can hold significant temperature gradients, which explains the superheating phase function of the coma of comet 67P/Churyumov-Gerasimenko measured by the Rosetta/VIRTIS (Visible and Infrared Thermal Imaging Spectrometer) instrument.  

\section{Methods}
 In this section, we present our numerical framework for solving radiative and conductive heat transfer in a particulate random medium enclosed by an arbitrarily shaped boundary. We neglect heat convection and assume that external radiation and thermal fluctuations are the only heat sources. We also assume that the radiative heat transport is much faster than the conductive heat transport. Thus, we consider the radiative part as instantaneous time harmonic radiation.  
 
\subsection{Energy equation}
Let us consider a finite body $\Omega \in \mathbb{R}^3$ characterized by the position $\bb r$, dependent density $\rho$, specific heat capacity $c_p$, and conductivity $\kappa$ in free space. Our goal is to solve the time $t$ dependent temperature $T$ under the influence of the external radiation source. The energy balance equation in $\Omega$ for the combined conductive and radiative heat transfer reads as 
\begin{equation}
\label{eq_energy1}
\rho c_{\rm p} \frac{\partial T}{\partial t} - \nabla \cdot \kappa\nabla T = -\nabla \cdot \bb q_{\rm r} \,\,\, \text{in} \, \Omega
\end{equation}
 with the boundary condition
\begin{equation}
  \bb n \cdot \kappa \nabla T = \bb n \cdot \bb q_{\rm r} \,\,\, \text{on} \, \partial \Omega
  \label{eq_energy2}
,\end{equation}
where $\bb q_{ \rm r}$ denotes the total radiative flux, $\partial \Omega$ is the boundary of $\Omega$, and $\bb n$ is the inward pointing normal vector on $\partial\Omega$. 

\subsection{Finite-element solution for the energy equation}

To solve the energy balance equation (\ref{eq_energy1}) and (\ref{eq_energy2}), we apply the standard finite-element method to discretize the spatial component of the energy equation. Taking the inner product with the testing functions $w$ and integrating by parts, the weak formulation reads as follows:

 Find $T \in H^1(\Omega)$ such that
\begin{equation}
\label{eq_weak}
\int_\Omega w \rho c_{\rm p} \frac{\partial T}{\partial t} \,\text{d}V  + \int_\Omega \nabla w \cdot \kappa \nabla T \,\text{d}V = -\int_\Omega w \nabla \cdot \bb q_{\rm r}\,\text{d}V 
\end{equation}
is valid for all $w \in H^1(\Omega)$. Here $H^1(\Omega)$ denotes the space of square integrable functions whose derivatives are also square integrable. We note that   equation (\ref{eq_weak}) automatically satisfies the Neumann boundary condition (\ref{eq_energy2}).

Next we discretize the time derivative using the finite-difference formula 
\begin{equation}
\label{eq_dtime}
\frac{\partial T}{\partial t} \approx \frac{1}{\tau}(T_{t+1} - T_{t})    
,\end{equation}
where the subscript $t$ indicates the time step and $\tau$ is its size. Substituting (\ref{eq_dtime}) into (\ref{eq_weak}) and using the implicit backward Euler time integration scheme, we obtain 
\begin{eqnarray}
&&\rho c_{\rm p} \tau^{-1}(\int_\Omega w T_{t+1} \,\text{d}V - \int_\Omega w T_{t} \,\text{d}V ) + \int_\Omega \nabla w \cdot \kappa \nabla T_{t+1} \,\text{d}V  = \nonumber \\
&&- \int_\Omega w (\nabla \cdot \bb q_{\rm r})_{t+1}.
\end{eqnarray}

We discretize the domain $\Omega$ with linear tetrahedral elements and expand the unknown temperature $T$ into the linear nodal basis functions $u^m$ as $T\approx \sum_{m} x^m u^m$, where $x^m$ are the unknown coefficients. Using Galerkin's method, in which the testing functions and basis functions are identical $w^m = u^m$ , we can write the mass and stiffness matrices as
\begin{equation}
M = \rho c_{\rm p} \int_\Omega w^n u^m\,\text{d}V,    
\end{equation}
\begin{equation}
S = \int_\Omega \nabla w^n \cdot \kappa \nabla u^m\,\text{d}V    
,\end{equation}
and the force vector as
\begin{equation}
\label{eq_force}
F = -\int_\Omega w^n \nabla \cdot \bb q_{\rm r}\,\text{d}V.     
\end{equation}
Finally, we arrive at the expression for the temperature coefficient vector $x$ given by
\begin{equation}
\label{eq_diff}
x_{t+1} = (M + \tau S)^{-1} (Mx_t + \tau Fx_{t+1}).    
\end{equation}
The force vector is strongly nonlinear with respect to temperature ($F\sim T^4$). Hence, estimating $\tau Fx_{t+1}$ with $\tau Fx_{t}$ would require using a very small time step $\tau$ to reach a converging numerical solution. Therefore, we solve  equation (\ref{eq_diff}) for $x_{t+1}$ by using an iterative method. In the next section we discuss how to compute the force vector $F$ in a densly packed particulate random medium enclosed by an arbitrarily shaped boundary $\partial \Omega$ with the Monte Carlo radiative transfer approach.

\subsection{Radiative transfer solution} 

To calculate the absorbed and thermally emitted power (i.e., the force vector $F$ in Eq. (\ref{eq_diff})), we employ an approximate version of the recently introduced radiative transfer with the reciprocal transactions (R$^2$T$^2$) method for electromagnetic scattering and propagation in dense discrete random media \citep{Muinonen2018, Markkanen2018, Vaisanen2019}. R$^2$T$^2$ is based on the Monte Carlo ray tracing of the order-of-scattering interactions   in which the fundamental scatterers are volume elements containing a large number of small grains. Incoherent electromagnetic interactions within the volume elements and the order-of-scattering interactions between the volume elements are computed via the exact electromagnetic solver, namely the fast superposition T-matrix method (FaSTMM) \citep{Markkanen2017}. 

Here we use the far-field approximation to compute the order-of-scattering interactions. Yet, the interactions within the volume elements are computed exactly. We also neglect the reciprocal scattering diagrams (i.e., those responsible for the coherent backscattering effect). This allows us to include the coherent field contribution in terms of reflections and refractions on the surfaces of a particle using geometric optics \citep{Markkanen2018_2, Jove2019}. 

Computations proceed as follows. First, we compute scattering properties for volume elements, as described in Section \ref{sec:sca_pro}. This step is done separately for each wavelength and for each material region, but it does not depend on the shape and size of the entire particle. Second, the computed scattering properties from the first step are used as input parameters in the combined Monte Carlo geometric optics radiative transfer solver, as discussed in Section \ref{sec:RT}. This step is computationally independent of the first step for different particle shapes and sizes.

\subsubsection{Scattering properties of volume elements}
\label{sec:sca_pro}
Input parameters for our radiative transfer algorithm are the incoherent scattering phase function $M_{11}^{\rm ic}$, incoherent albedo $\omega$, incoherent mean free path $\ell$, coherent effective refractive index $m_{\rm eff}$, and absorption cross section of the ensemble averaged volume element $C_{\rm abs}$. 

We compute these parameters for spherical ensemble-averaged volume elements that contain a large number of small spherical monomers. The radius of the volume element $R_0$ should be large enough to capture the statistics of the random medium. First, we generate a sample medium into a periodic box by randomly depositing monomers into the box until the target volumetric filling factor $v$ is reached. Second, we draw a sample spherical volume element from the box, as explained in detail by \cite{Vaisanen2019}. 
Third, we apply FaSTMM to compute the Jones matrix $S_i$ of such a volume element $i$. Repeating the process $N$ times ($N = 512$ is used in this work), we compute the coherent Jones scattering matrix as
\begin{equation}
    S^{\rm c} = \frac{1}{N} \sum_{i=1}^N  S_i,
\end{equation}
and the incoherent Jones scattering matrix for the $i$th state as
\begin{equation}
    S^{\rm ic}_i = S_i - S^{\rm c}.
\end{equation}
Averaging the square of the incoherent Jones matrix over the ensemble of all $N$ sample volume elements, we obtain the ensemble-averaged one-one element of the Mueller matrix,  composed of the incoherent scattering phase function 
\begin{equation}
    M_{11}^{\rm ic} = \frac{1}{N} \sum_i^N |S^{\rm ic}_{i}|^2
\end{equation}
and the incoherent scattering cross section
\begin{equation}
\label{eq_csca_ic}
    C_{\rm sca}^{\rm ic} = \int_{S^2} M_{11}^{\rm ic} \,\text{d}S
,\end{equation}
where $S^2$ is the surface of the unit sphere.
The incoherent albedo is defined as 
\begin{equation}
\omega= \frac{C_{\rm sca}^{\rm ic}}{C_{\rm sca}^{\rm ic} + C_{\rm abs}}
,\end{equation}
where $C_{\rm abs}$ is the ensemble-average absorption cross section, and 
the incoherent mean free path length as
\begin{equation}
    \ell = \frac{V}{C^{\rm ic}_{\rm sca} + C_{\rm abs}}
,\end{equation}
in which $V$ is the volume of the volume element.

Finally, we solve the coherent effective medium parameter $m_{\rm eff}$. This is done by matching the coherent scattering cross section $C^{\rm c}_{\rm sca}$ to the $C_{\rm sca}$ of the equal-sized sphere using the Mie solution. $C^{\rm c}_{\rm sca}$ is defined as in (\ref{eq_csca_ic}), but the incoherent Mueller matrix is replaced with the coherent matrix. 

\subsubsection{Monte Carlo ray tracing}
\label{sec:RT}
We trace the rays in the same tetrahedral mesh as is used in the FEM. This makes it trivial to couple the radiative transfer solution (i.e., $\nabla \cdot \bb q_{\rm r}$)  to the FEM solution. We decompose the divergence of the total flux into two parts $\nabla \cdot \bb q_{\rm r} = q_{\rm sol} + q_{\rm em}$ in which $q_{\rm sol}$ is the absorption rate due to the solar radiation, and $q_{\rm em}$ is the absorption plus emission rate due to the thermally fluctuating sources.

To compute $q_{\rm sol}$, we launch $N^e_{\rm ray}$ rays for a given wavelength band  $\Delta_\lambda$ from outside $\Omega$. The power of each incident ray is given by
\begin{equation}
E^e = \int_{\lambda_1}^{\lambda_2} \frac{P(\lambda) G }{N^e_{\rm ray}} \,\text{d}\lambda     
,\end{equation}
where $P(\lambda)$ is the flux density of the incident radiation and $G$ is the geometric cross section of the particle. Each ray has a specific position $\bb r$ and propagation direction $\hat{\bb k}$ depending on the source. 

For thermally emitted radiation, to compute $q_{\rm em}$, $N^i_{\rm ray}$ rays are launched from each tetrahedron $C_t$ with a random position and direction. The emitted power per ray for a given wavelength is given by   
\begin{equation}
    E^i_{C_t} = \frac{4 N_{\rm d}}{N^i_{\rm ray}}\int_{\lambda_1}^{\lambda_2} \pi B(\lambda, T_{c_t}, m_{\rm eff}) C_{\rm abs}\,\text{d}\lambda
,\end{equation}
where $T_{c_t}$ is the average temperature of the tetrahedron $C_t$ and $N_{\rm d}$ is the number density of the volume elements in the tetrahedron $C_t$. The Planck function in a medium with the refractive index $m$ written in terms of the free space wavelength $\lambda_0$ reads as
\begin{equation}
    B_\lambda(\lambda_0,T,m) = \frac{2hc^2 Re\{m\}^2}{\lambda_0^5} \frac{1}{\exp( \frac{hc}{\lambda_0 k_{\rm b} T})-1},
\end{equation}
where $h$ is the Planck constant, $c$ is the speed of light in vacuum, and $k_{\rm b}$ is the Stefan-Boltzmann constant.

Once we have the initial position $\bb r$, direction $\hat{\bb k}$, and power of the ray $E$, we start tracing. If the ray hits the boundary $\partial\Omega$, it reflects and refracts according to Snel's law and the power is updated from Fresnel's coefficients which are calculated using the effective refractive index $m_{\rm eff}$. The position and direction of the reflected and refracted rays are updated. 

If the ray is in $\Omega$, the distance to the next scattering event is generated as
\begin{equation}
    d = -\ell \ln \Phi
,\end{equation}
where $\Phi$ is a uniform random number within $[0,1[$. The scattering event happens if the ray does not cross the boundary $\partial \Omega$. Absorbed power $E_{\rm abs} = E(1-\omega)$ in the scattering event is added into the total energy of the tetrahedron where the scattering event occurs, and the ray's power is updated ($E = E \omega$). A new propagation direction $\hat{\bb k}$ is drawn from the cumulative distribution function generated from the phase function of the ensemble averaged incoherent volume element $M_{11}^{\rm ic}$. Finally, if the ray does not scatter and it does not cross the boundary $\partial \Omega$ , intensity is collected and added to the total scattering phase function. This process is repeated until the ray's power has decreased under the predefined threshold.

The radiative transfer computation gives us $\int_{C_t} \nabla \cdot \bb q_{\rm r} \,\text{d}V$ for each tetrahedron, which is then used to compute the force vector $F$ using (\ref{eq_force}). In addition, since the escaped rays are recorded, the total scattering and thermal emission phase functions and cross sections are obtained for each wavelength.   

\section{Numerical results}

In this section, we present some numerical results. To validate the method, we compare our solution with the known approximate solution for small loosely packed grains computed by employing the effective medium approximation (EMA) and the Mie theory. Then, we apply the method to explain the superheating factor phase function measured for the coma of comet 67P/Churyumov-Gerasimenko.

\subsection{Comparison to the effective medium approximation}

\begin{figure}[htb]
\includegraphics[width=0.5\textwidth]{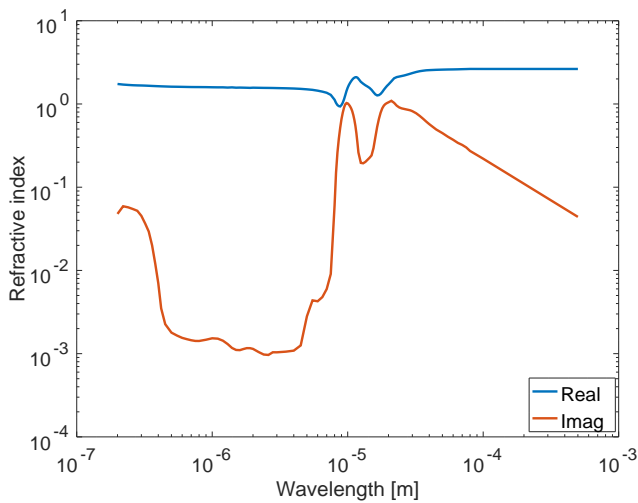}
\caption{Real and imaginary parts of the complex refractive index for magnesium-iron silicate.}
\label{fig1}
\end{figure}

Let us first consider a spherical medium consisting of small spherical silicate grains of radius $r = \SI{20}{\nano\metre}$. The grains are randomly positioned in a spherical domain $\Omega$ with radius $R$. The volumetric filling factor is $\nu = 0.05$. The wavelength dependent refractive index for magnesium-iron silicate,   from \cite{Dorschner1995}, is plotted in Fig. \ref{fig1}. Here we assume that the refractive index is independent of temperature in the studied temperature range.

The external radiation source is the Sun, which is assumed to radiate as a perfect blackbody of temperature $T_\odot = 5777$K. The solar flux $P$ in the given wavelength band ($\lambda_1 - \lambda_2$) at the distance $\Delta = 1.3$AU from the Sun is given by 
\begin{equation}
 P = (\frac{r_\odot}{\Delta})^2 \int_{\lambda_1}^{\lambda_2} \pi B_\lambda(\lambda,T_\odot) \,\text{d}\lambda    
,\end{equation}
where $r_\odot$ is the radius of the Sun. 

Since the grain size is small, we can calculate a reference solution at the equilibrium temperature using the steady-state energy balance equation, EMA, and the Mie theory. To compute the absorption cross section $C_{\rm abs}$, we use the Maxwell-Garnett mixing rule \citep{Sihvola2000} and the Mie theory. For an isothermal spherical particle, the steady-state energy balance equation reads as
\begin{equation}
    (\frac{r_\odot}{\Delta})^2 \int_{0}^{\infty} \pi B_\lambda(\lambda,T_\odot) C_{\rm abs}(\lambda) \,\text{d}\lambda
  = 4\int_0^{\infty} \pi B_\lambda(\lambda,T) C_{\rm abs}(\lambda)\,\text{d}\lambda.
  \label{eq_ss}
\end{equation}

\begin{figure}[htb]
\includegraphics[width=0.5\textwidth]{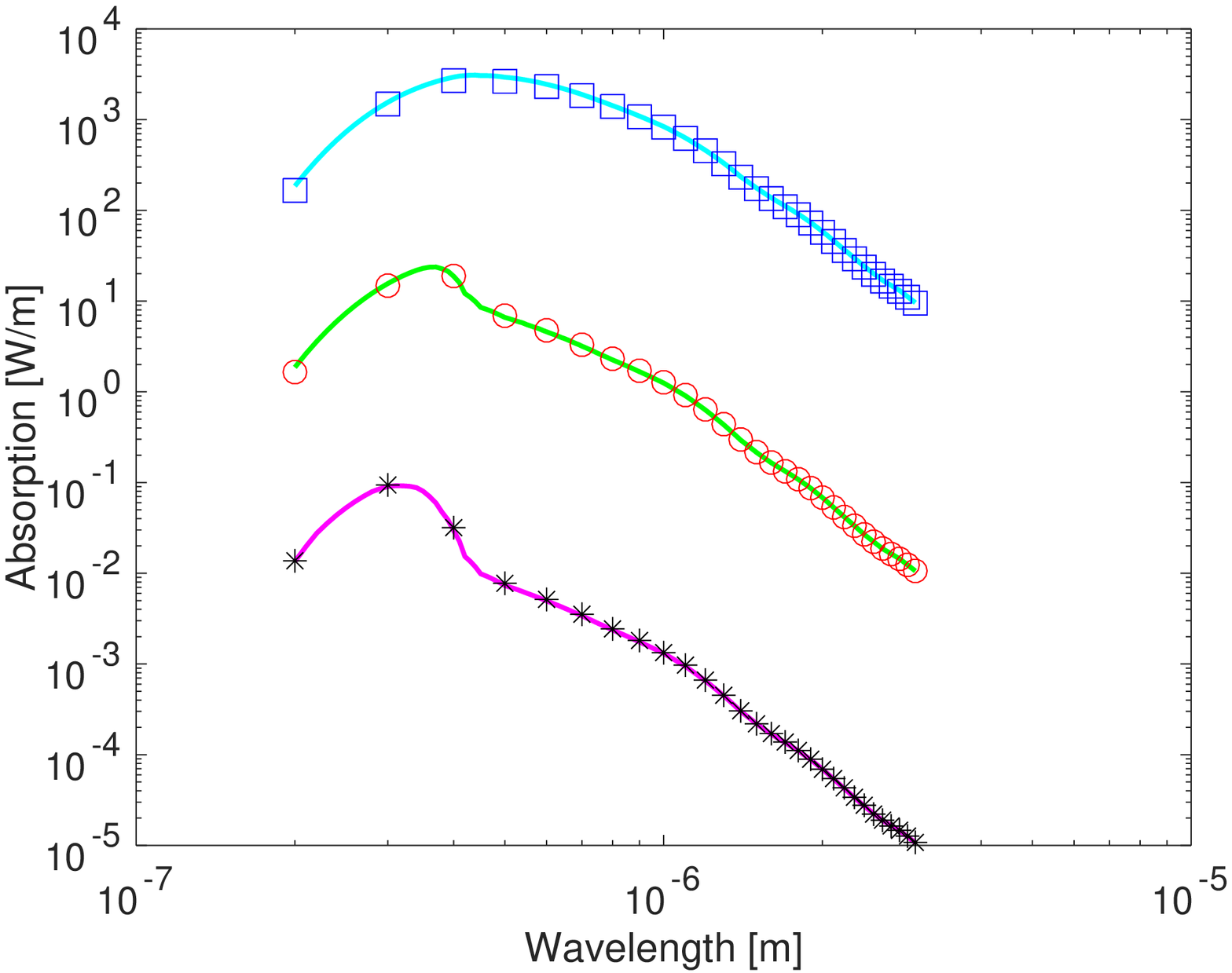}
\includegraphics[width=0.5\textwidth]{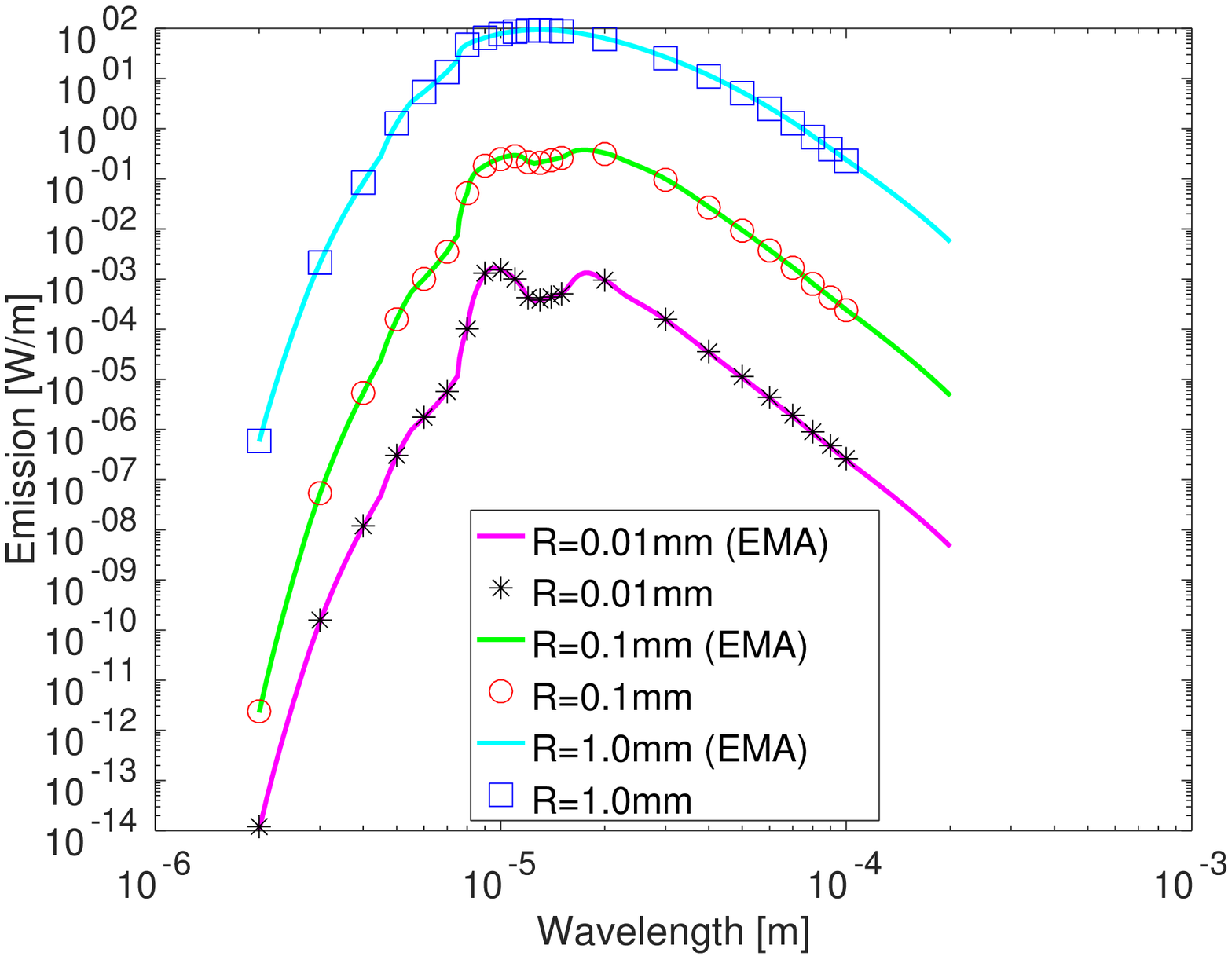}
\caption{Total absorbed power of solar radiation (top) and thermally emitted power (bottom) by particles with radii \SI{10}{\micro\metre},  \SI{100}{\micro\metre,} and \SI{1.0}{\milli\metre} as a function of wavelength computed by the proposed method and the effective medium approximation (EMA) are shown. Grain size is \SI{20}{\nano\metre} and the volumetric filling factor $v=0.05$.}
\label{fig2}
\end{figure}

To compare the proposed method to the steady-state  analysis with EMA (\ref{eq_ss}), we ran the transient simulation until the equilibrium was reached. Figure\ \ref{fig2} shows the total absorbed solar radiation and thermally emitted power as a function of wavelength calculated by the proposed method and EMA for three different particle sizes $R=0.01, 0.1$, and $1.0$\,mm. We observed excellent agreement between the two methods; however, we  expect a decrease in the accuracy of our method as the particle size becomes smaller. This is evident as we   used geometric optics to approximate the coherent reflections and refractions on the surface of a particle.   

\begin{figure}[htb]
\includegraphics[width=0.5\textwidth]{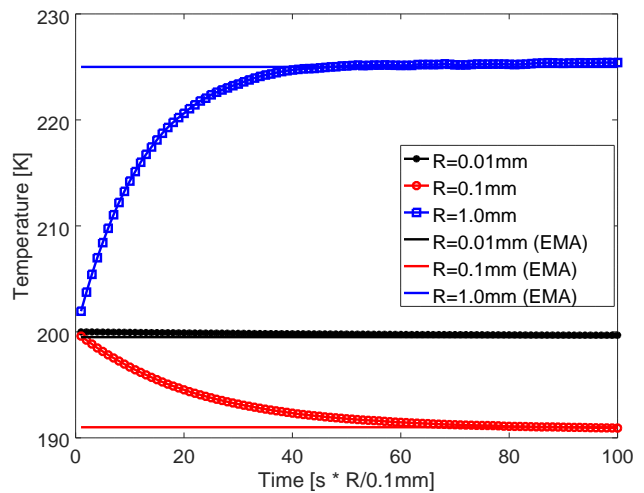}
\caption{Temperature time evolution for porous ($v=0.05$) aggregates of 20\,nm silicate grains, with initial temperature $T_0=200$K. The unit on the time axis is scaled by the radii of the particles $R$.}
\label{fig3}
\end{figure}

In the computations the volume element radius is $r_0 = \SI{0.5}{\micro\metre}$ and the time step is $\tau = \SI{0.1}{\second}$, $\tau = \SI{1.0}{\second}$, and $\tau = \SI{10}{\second}$ for \SI{0.01}{\milli\metre}, \SI{0.1}{\milli\metre}, and \SI{1.0}{\milli\metre} particles, respectively. We also assumed high thermal conductivity $\kappa = 100$ W/m/K to remove possible temperature gradients inside the body. Although it is  not a realistic assumption for real dust, it serves as a validation  by simplifying the problem. It is worth noting that the density $\rho = 1000$ kg/m$^3$ and the specific heat capacity $c_p = 500$ J/kg/K do not affect the equilibrium temperature as $\frac{\partial T}{\partial t}=0$ in (\ref{eq_energy1}) for the steady-state solution; they only affect the time it takes to reach equilibrium. Figure \ref{fig3} shows the temperature evolution of the particles when the initial temperature $T_0=200$K. We see that the transient temperature reaches the equilibrium temperature computed by using the steady-state approach with EMA.  

\begin{figure}[htb]
\includegraphics[width=0.5\textwidth]{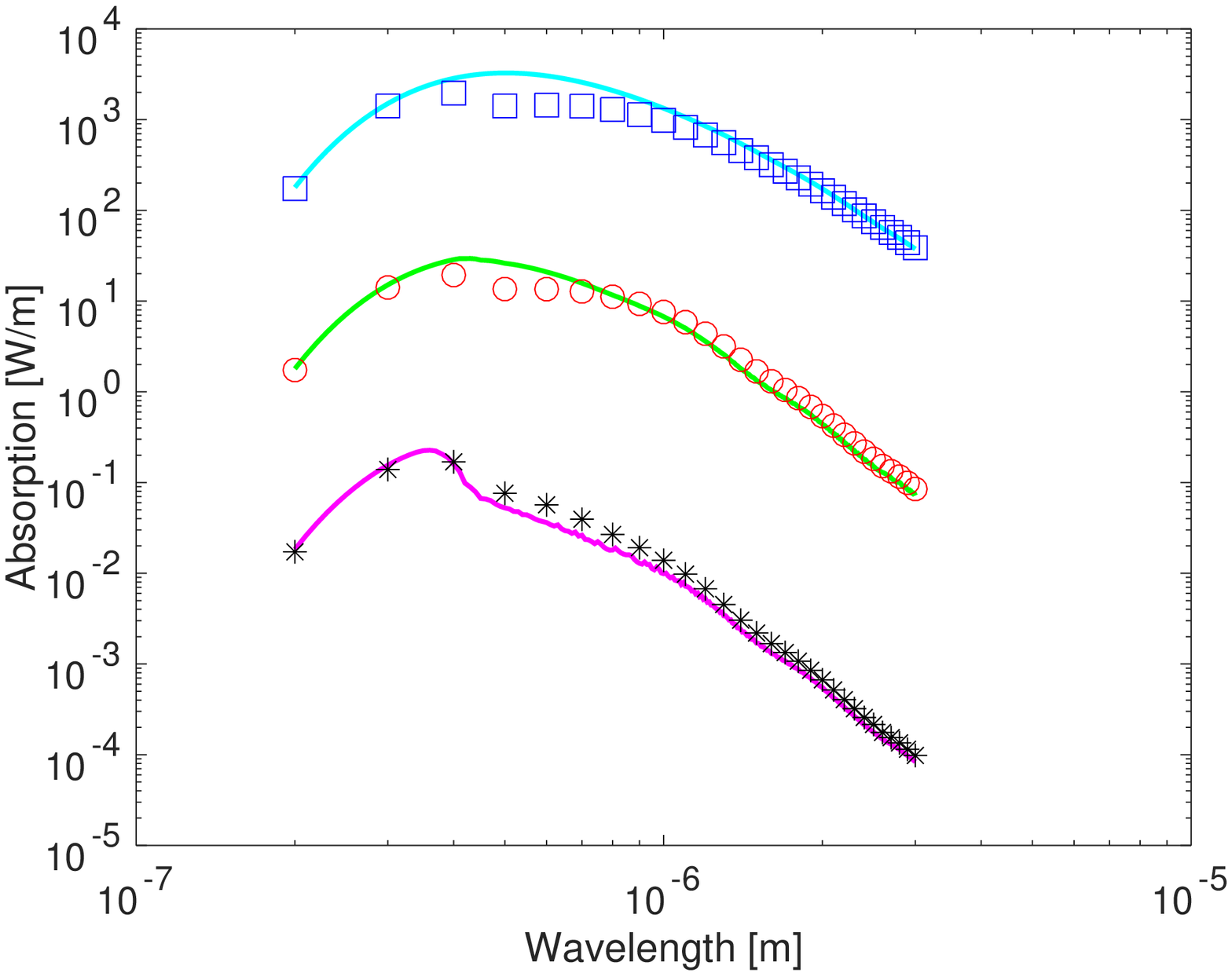}
\includegraphics[width=0.5\textwidth]{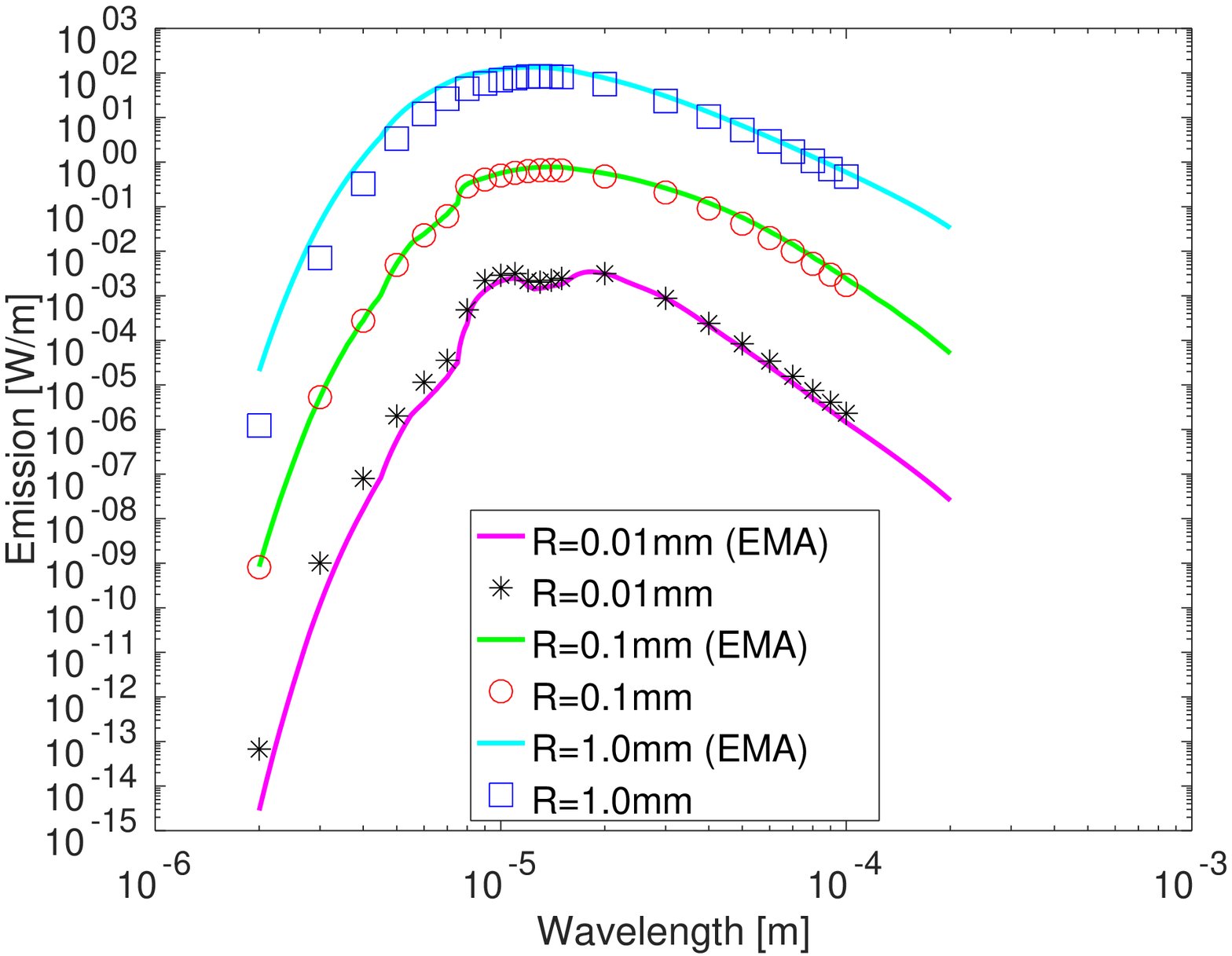}
\caption{As in Fig. \ref{fig2}, but for  grain size of \SI{100}{\nano\metre} and  filling factor $v=0.3$. }
\label{fig4}
\end{figure}

\begin{figure}[htb]
\includegraphics[width=0.5\textwidth]{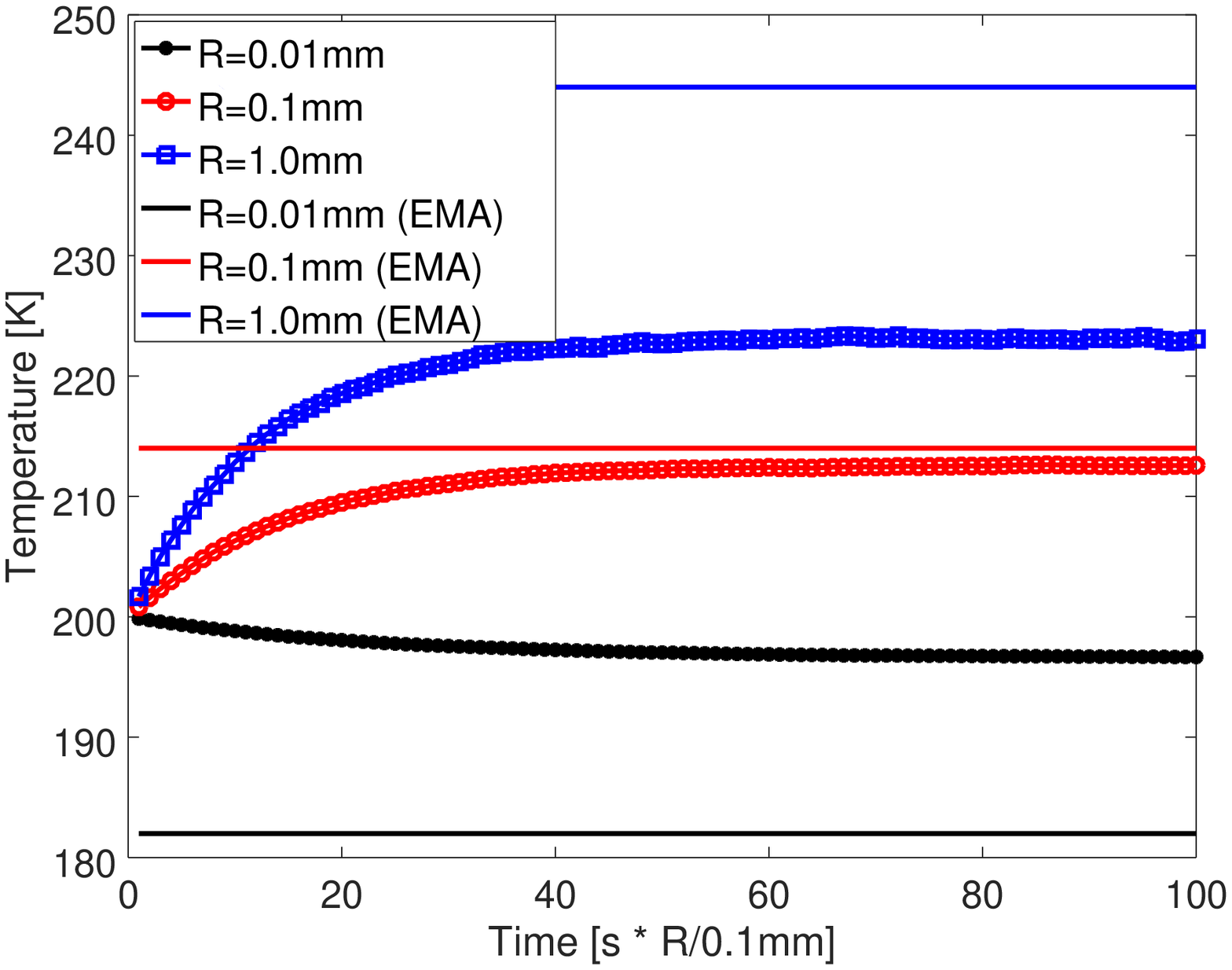}
\caption{As in Fig. \ref{fig3}, but for  grain size of \SI{100}{\nano\metre} and  filling factor $v=0.3$.}
\label{fig5}
\end{figure}

Next we studied larger grains with $r = 100$\,nm and increased the filling factor to $v = 0.3$. These values are commonly used to model cometary dust \citep{Kolokolova2004}. It is evident from Fig.\ \ref{fig4} and  Fig.\ \ref{fig5}  that EMA leads to a different absorption rate, which consequently affects the equilibrium temperature and thermal emission;  the EMA is no longer a valid approximation at shorter wavelengths as the grain size is approximately the same as the wavelength.  Thus, care should be taken when applying EMAs to interpret infrared observations of cometary dust particles consisting of 100\,nm-sized grains.    

\subsection{Application to the coma of 67P}

We recently introduced a particle model that explains the scattering phase functions of the coma of comet 67P/Chruyumov-Gerasimenko measured by the Rosetta/OSIRIS (Optical, Spectrocopic and Infrared Remote Imaging System) instrument \citep{Markkanen2018_2}. The particle model consists of aggregated submicrometer-sized organic grains and micrometer-sized silicate grains. The shapes of the aggregates are     Gaussian random spheres. It is thus interesting to study whether the same model can also explain the thermal infrared observations of the same comet. Since there is no refractive index available for cometary organic material over a wide wavelength range, we used the refractive index of carbonaceous dust analogues \citep{Jaeger1998}. For silicates, we used the refractive index presented in Fig. \ref{fig1} that corresponds to magnesium-iron silicate mineral \citep{Dorschner1995}. 

We concentrate on the superheating factor phase function reported by \cite{Dominique2019} for the coma of the comet 67P. The superheating factor is defined as the ratio of the color temperature $T_{\rm c}$ to the equilibrium temperature of an ideal blackbody $T_{\rm b} = 278/\Delta^2$[K] (i.e., $S = T_{\rm c}/T_{\rm b}$). To obtain the color temperature $T_{\rm c}$ for a given phase angle, \cite{Dominique2019} fitted the Planck function to the Rosetta/VIRTIS near-infrared spectra ($3-\SI{5}{\micro\metre}$). They found that the superheating factor phase function has a phase dependence indicating that the particles can hold temperature gradients if the particles are assumed to be randomly oriented.

\begin{figure}[htb]
\includegraphics[width=0.5\textwidth]{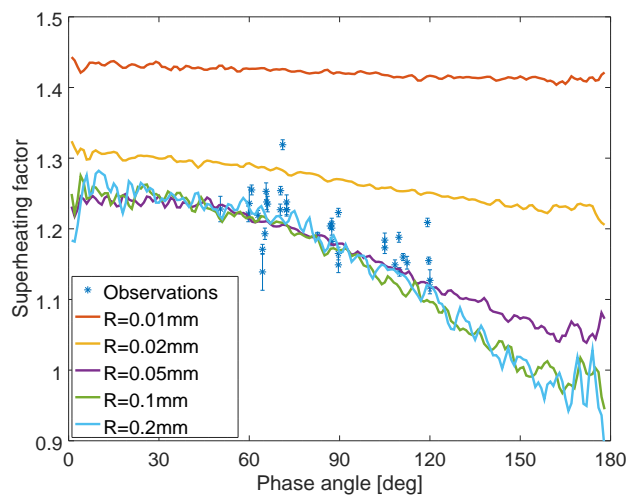}
\caption{Computed superheating factors as a function of phase angle for different particle sizes. The heat conduction coefficient is $\kappa = 0.00025$\,W/m/K. Also shown are the measured values.}
\label{fig6}
\end{figure}

\begin{figure}[htb]
\includegraphics[width=0.5\textwidth]{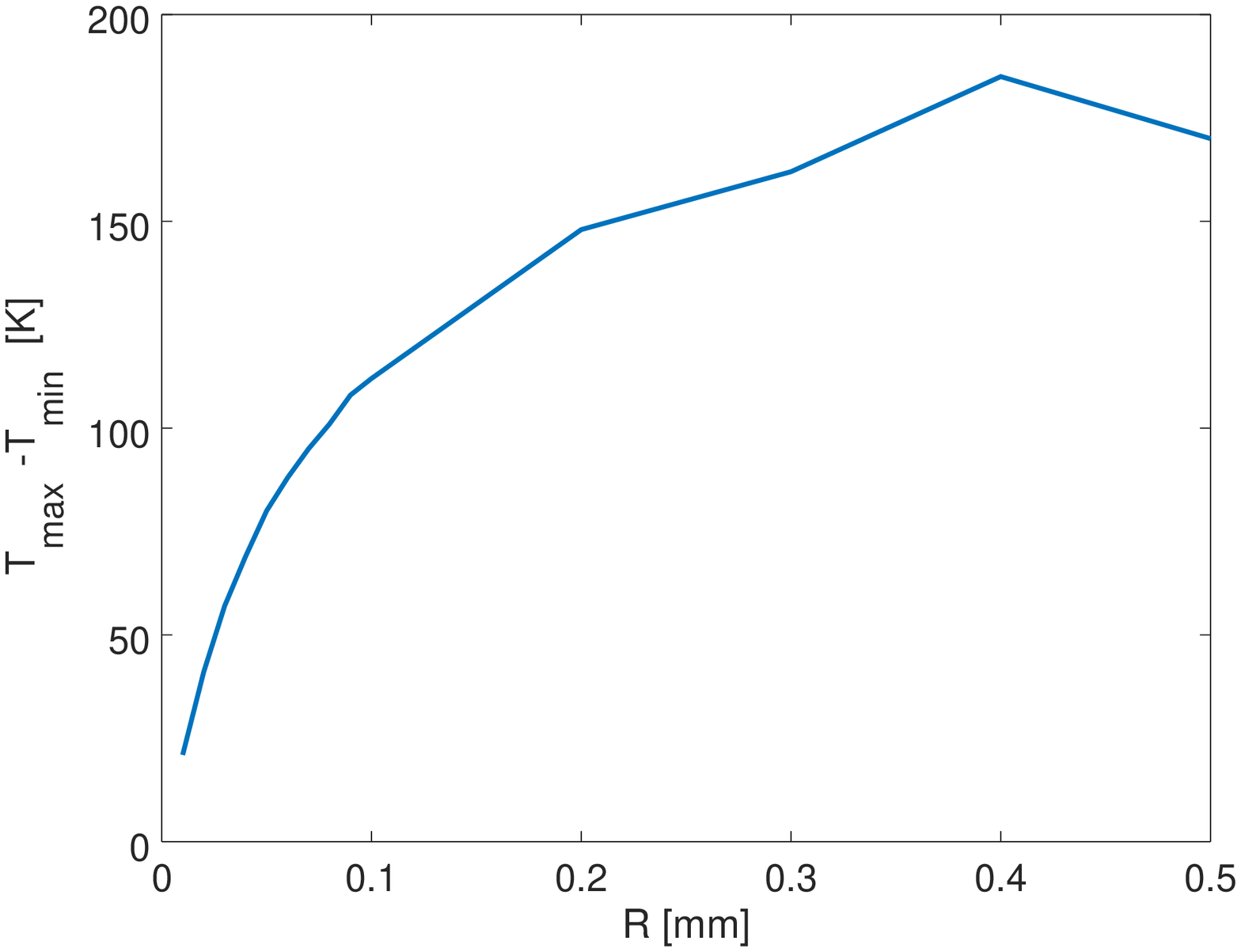}
\caption{Difference between the maximum and minimum temperature inside the particle as a function of particle radius for an example particle.}
\label{fig7}
\end{figure}

Computed superheating factors for different particle sizes together with the measured values (Fig. 8 in \cite{Dominique2019}) as a function of phase angle are presented in Fig. \ref{fig6}. We see that only large particles can sustain  temperature gradients resulting in the phase functions with significant slopes. The difference between the maximum and minimum temperature inside the particle with respect to the particle size is plotted in Fig. \ref{fig7}. It is evident that in order to model the measured superheating factor phase function slopes, the dust particles must sustain significant temperature gradients. Large temperature gradients, in turn, may have interesting effects on the dust dynamics in the inner coma via rocket forces caused by the asymmetrical sublimation of ices. Thermal break-down of particles may also play a role as the dust is transported from the inner to outer coma. Nevertheless, these interesting topics are beyond of the scope of this paper and are left for future research.

In our computations, the heat conduction coefficient is assumed to be $\kappa = 0.00025$\,W/m/K, which corresponds to a realistic value for a porous dust aggregate \citep{Krause2011, Arakawa2017, Sakatani2017}. The heat conduction coefficient depends on porosity, grain size, and composition. Increasing the heat conduction coefficient   flattens the superheating phase functions. Decreasing $\kappa$ increases the slope until the radiative heat transport becomes more dominant. We   also assumed that the angular velocities of the particles in the direction perpendicular to the Sun are slow ($\omega = \frac{2\pi}{50} \frac{0.1\text{mm}}{R}$ rad/s), hence the contribution of rotation to the temperature gradients is negligible. The slow rotation rate was introduced in order to speed up numerical convergence. Higher angular velocities decrease  temperature gradients and flatten  the superheating factor phase functions. However, it is important to note that if the rotation axis points towards the Sun, it does not affect the temperature gradients.     

\begin{figure}[htb]
\includegraphics[width=0.5\textwidth]{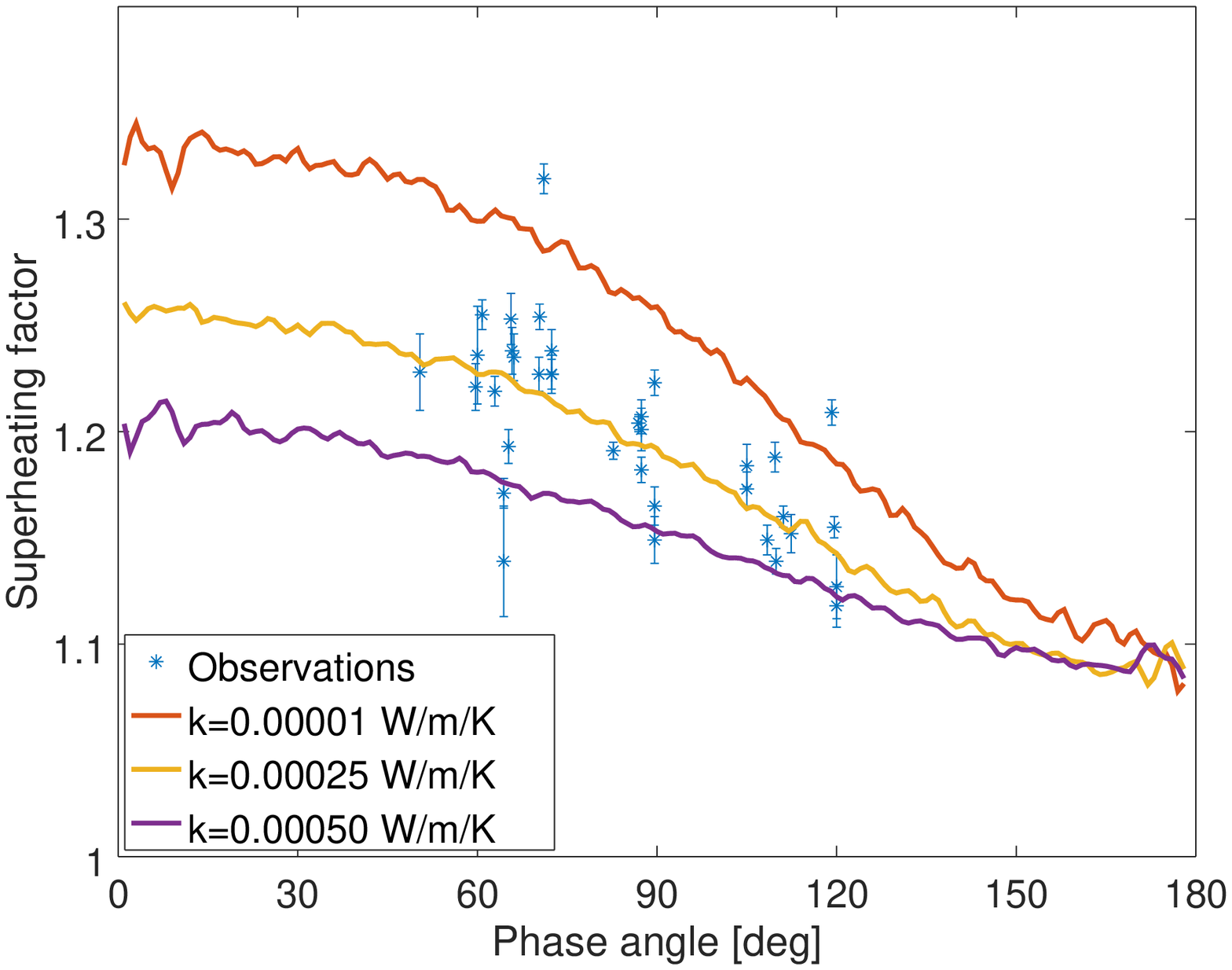}
\caption{Modeled superheating factor phase function for different conductive heat transfer coefficients $\kappa$  averaged over the differential power-law size distribution of index -3. The minimum cutoff is \SI{10}{\micro\metre} and the maximum \SI{100}{\micro\metre}.}
\label{fig8}
\end{figure}

Finally, we averaged the modeled superheating phase functions   over a differential power-law size distribution of index -3. The averaging results in a superheating factor phase function that is consistent with the measured value if the conductive heat coefficient $\kappa =  0.00025$\,W/m/K, as demonstrated in Fig \ref{fig8}. Such  a low $\kappa$ value suggests that the particles must be porous. The superheating factor phase function also indicates that the dust particles in the inner coma must be large. The dominating particle size range is around 10-\SI{100}{\micro\metre}. Smaller particles would show flatter superheating factor phase functions with higher absolute values whereas larger particles would result in a steeper slope and smaller absolute values. The size range is also consistent with other dust models explaining the intensity phase functions measured by Rosetta/OSIRIS \citep{Moreno2018, Markkanen2018_2}. 
 
 \section{Conclusions}
 
 We  presented a synoptic numerical solution for scattering, absorption, and thermal emission by large aggregated dust particles consisting of submicrometer-sized grains. The numerical solution is self-consistent and allows for analysis of multi-instrument data sets. We  solved the transient energy equation using the finite-element method in which the radiative heat transfer part is included as an additional forcing term.  We developed a new algorithm, based on the Monte Carlo radiative transfer with reciprocal transactions framework, to compute the forcing term in the dense discrete random medium enclosed by an arbitrarily shaped surface. The method explicitly computes the radiative heat transfer part in contrast to the standard thermal models in cometary science where it is included in the heat conduction coefficient by assuming that heat transport is one dimensional. Thus, the developed method provides   a full three-dimensional solution for radiative heat transfer problems.  
 
 We compared the solution of the proposed method to that obtained by employing the effective medium approximation and Mie theory. The solutions match if the grains are small and loosely packed, but deviate for grain sizes and packing densities that are typically assumed to form cometary dust particles. This implies that using EMAs to interpret thermal infrared observations of cometary dust may lead to unreliable results. Finally, we  showed that particles $10-\SI{100}{\micro\metre}$ in size  can reproduce the superheating factor phase function of the coma of comet 67P if the dust particles hold significant temperature gradients. This means that the heat conduction coefficient must be very low, hence particles must be porous. The presented analysis also suggests that the visible phase functions measured by Rosetta/OSIRIS and the superheating phase function from Rosetta/VIRTIS are consistent with each other.


\begin{acknowledgements}
This work has been funded by the ERC Starting Grant No. 757390 Comet and Asteroid Re-Shaping through Activity (CAstRA). Computational resources have been provided by Gesellschaft f\"{u}r Wissenschaftliche Datenverarbeitung mbH G\"{o}ttingen (GWDG).
\end{acknowledgements}

   \bibliographystyle{aa} 
   \bibliography{refs.bib} 

\end{document}